\documentclass[10pt,a4paper,twocolumn]{article}

\usepackage[a4paper,top=1.8cm,bottom=1.8cm,left=1.7cm,right=1.7cm]{geometry}
\usepackage{microtype}

\usepackage{amsmath,amssymb,amsfonts}
\usepackage{bm}
\usepackage{mathtools}

\usepackage{graphicx}
\usepackage{booktabs}
\usepackage{array}
\usepackage{multirow}
\usepackage{caption}
\usepackage{subcaption}
\usepackage{slashed,stmaryrd}
\usepackage{authblk}

\usepackage[numbers,sort&compress]{natbib}
\usepackage[colorlinks=true,citecolor=blue,linkcolor=blue,urlcolor=blue]{hyperref}
\usepackage[nameinlink,noabbrev]{cleveref}

\usepackage{titlesec}
\titleformat{\section}{\normalsize\bfseries}{\thesection.}{0.45em}{}
\titleformat{\subsection}{\small\bfseries}{\thesubsection.}{0.45em}{}
\titlespacing*{\section}{0pt}{0.8em}{0.35em}
\titlespacing*{\subsection}{0pt}{0.6em}{0.25em}

\newcommand{\keywords}[1]{%
	\vspace{0.45em}\noindent\textbf{Keywords:} #1\par
}

\title{An interrelation of three neutrino mixing angles}

\author[1,2,3]{Zhi-zhong Xing\thanks{ORCID: 0000-0002-9652-4989; Email: xingzz@ihep.ac.cn}}

\affil[1]{Institute of High Energy Physics, Chinese Academy of Sciences, Beijing 100049, China}
\affil[2]{School of Physical Sciences, University of Chinese Academy of Sciences, Beijing 100049, China}
\affil[3]{Center of High Energy Physics, Peking University, Beijing 100871, China}

\date{}

\begin{document}
	
\maketitle
	
\begin{abstract}
We conjecture a novel interrelation of three neutrino mixing angles:
$\theta^{}_{23} = \left(\theta^{}_{12} + \theta^{}_{13}\right)/2 + \pi/8$,
which predicts $\theta^{}_{23} = 43.46^\circ \pm 0.26^\circ$ with
inputs of the latest JUNO and Daya Bay precision data. A simple ansatz
is proposed to derive this phenomenologically favored and experimentally 
testable angle relation, and a new expansion of the lepton flavor mixing 
matrix in powers of a single small parameter is presented as a by-product.	
\end{abstract}
	
\keywords{neutrino oscillations; tribimaximal flavor mixing; JUNO
measurements}
	
	
	
\def\thefootnote{\arabic{footnote}}
\setcounter{footnote}{0}
\setcounter{equation}{0}
\setcounter{figure}{0}

\section{Motivation}

The JUNO reactor antineutrino oscillation experiment has recently
reported its precision measurements of the solar neutrino oscillation
angle and two neutrino mass-squared differences in the
three-flavor scheme~\cite{JUNO:2025gmd,JUNO:2026tap}:
\begin{eqnarray}
&& \sin^2\theta^{}_{12} = 0.3036 \pm 0.0064 \; ,
\nonumber \\
&& \Delta m^2_{21} \equiv m^2_2 - m^2_1
= \left(7.388 \pm 0.078\right) \times 10^{-5} ~{\rm eV} \; ,
\hspace{0.34cm}
\nonumber \\
&& \Delta m^2_{31} \equiv m^2_3 - m^2_1
= 2.509^{+0.027}_{-0.025} \times 10^{-3} ~{\rm eV} \; ,
\label{1}
\end{eqnarray}
which favor the normal mass ordering $m^{}_1 < m^{}_2 < m^{}_3$
over the inverted one $m^{}_3 < m^{}_1 < m^{}_2$ at the
$2.1\sigma$ confidence level. The ratio $\Delta m^2_{21}/
\Delta m^2_{31} = 0.02945^{+0.00061}_{-0.00062}$ turns out to be
compatible with the Alikhanov relation
$\Delta m^2_{21}/\Delta m^2_{31} = \tan^4 (\pi/8)
= (\sqrt{2} - 1)^4 \simeq 0.02944$ to a remarkably good
degree of accuracy~\cite{Alikhanov:2026jmw,Alikhanov:2026bhi}.
The success of such an empirical relation for neutrino masses
poses a meaningful question to us: may there exist an
interrelation of three neutrino mixing angles $\theta^{}_{12}$,
$\theta^{}_{13}$ and $\theta^{}_{23}$ favored by current neutrino
oscillation data and implied by an underlying flavor symmetry?

\begin{figure}[t]
\vspace{0.2cm}
\begin{center}
\includegraphics[width=7.4cm]{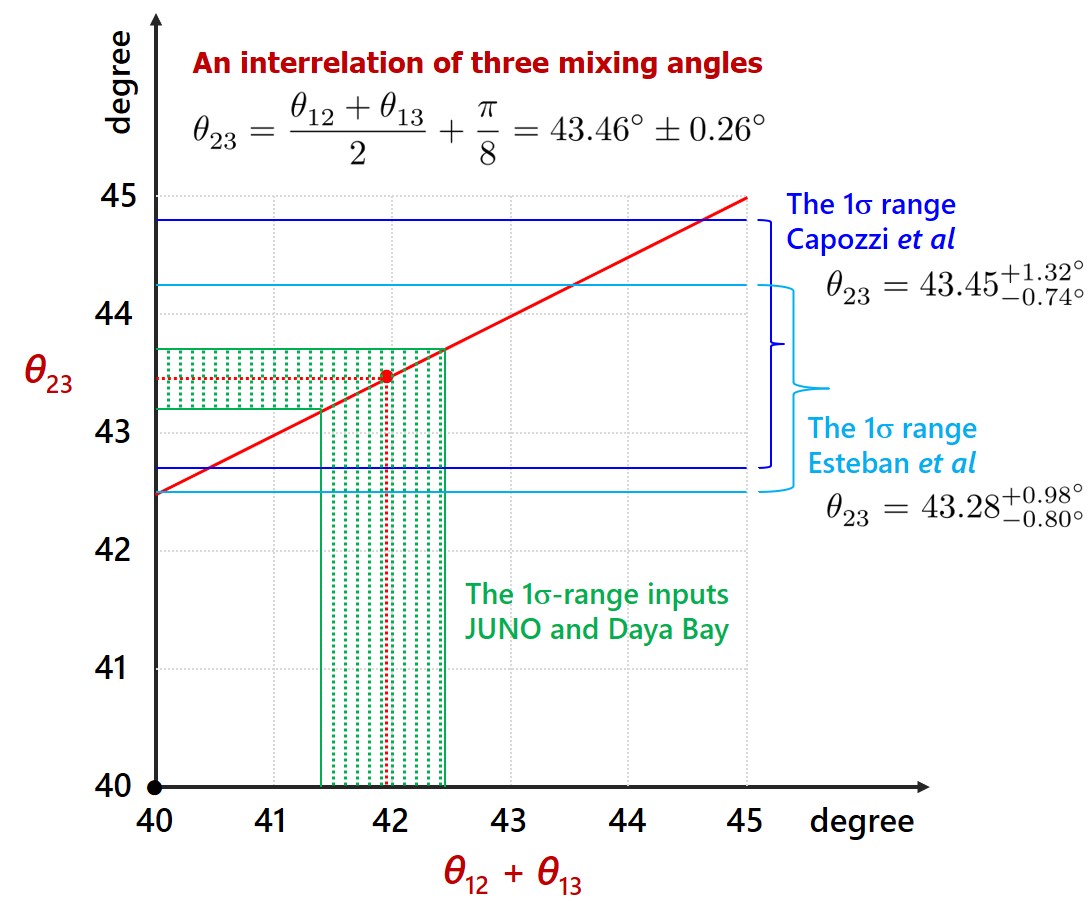}
\vspace{-0.2cm}
\caption{A numerical prediction for $\theta^{}_{23}$ with the JUNO and
Daya Bay inputs of $\theta^{}_{12}$ and $\theta^{}_{13}$, as compared
with the global-fit values of $\theta^{}_{23}$ reported by Capozzi
{\it et al}~\cite{Capozzi:2025wyn} and by Esteban
{\it et al}~\cite{Esteban:2024eli} at the $1\sigma$ confidence level.}
\label{fig1}
\end{center}
\end{figure}

The present letter aims to provide a partial but affirmative answer
to this question by conjecturing a novel angle relation of the form
\begin{eqnarray}
\theta^{}_{23} = \frac{\theta^{}_{12} + \theta^{}_{13}}{2} +
\frac{\pi}{8} \; ,
\label{2}
\end{eqnarray}
where $\theta^{}_{12} = 33.44^{+0.39^\circ}_{-0.40^\circ}$
is extracted from Eq.~(\ref{1}), and
$\theta^{}_{13} = 8.48^\circ \pm 0.12^\circ$ can be achieved
from the Daya Bay precision measurement
$\sin^2 2\theta^{}_{13} = 0.0851 \pm 0.0024$~\cite{DayaBay:2022orm}.
As a result, Eq.~(\ref{2}) predicts
$\theta^{}_{23} = 43.46^\circ \pm 0.26^\circ$, which is
consistent very well with the value $\theta^{}_{23} =
43.45^{+1.32^\circ}_{-0.74^\circ}$~\cite{Capozzi:2025wyn}
(or $43.28^{+0.98^\circ}_{-0.80^\circ}$~\cite{Esteban:2024eli})
that has been obtained from a global analysis of today's neutrino
oscillation data at the same $1\sigma$ confidence level
(see Fig.~\ref{fig1} for a numerical illustration).
We are therefore motivated to see how to make Eq.~(\ref{2})
reasonably hold, at least from a phenomenological point of view.

\section{The magic angle basis}

To derive the interrelation of three neutrino mixing angles
conjectured in Eq.~(\ref{2}), let us start from two magic constant
angles that are closely associated with the well-known
democratic~\cite{Fritzsch:1995dj,Fritzsch:1998xs,Fritzsch:1999ee}
and tribimaximal~\cite{Harrison:2002er,Xing:2002sw,He:2003rm} lepton
flavor mixing patterns:
\begin{eqnarray}
\theta^{\star}_{12} = \arctan(1/\sqrt{2}) \simeq 35.26^\circ \; ,
\quad \theta^{\star}_{23} = 45^\circ \; ,
\label{3}
\end{eqnarray}
which can be naturally obtained from a number of discrete flavor
symmetry models for massive
neutrinos~\cite{Xing:2019vks,Feruglio:2019ybq,Ding:2024ozt}.
It has long been noticed that the smallest lepton flavor mixing
angle $\theta^{}_{13}$ may originate from a proper correction to
the constant angle
\begin{eqnarray}
\theta^{\star}_{13} \equiv \theta^{\star}_{23} - \theta^{\star}_{12}
= \arctan\big(\sqrt{2} - 1\big)^2 \simeq 9.74^\circ ,
\label{4}
\end{eqnarray}
which simply stems from a shift from the democratic flavor mixing
pattern to the tribimaximal one~\cite{Xing:2010pn}. A combination
of Eqs.~(\ref{3}) and (\ref{4}) leads us to the relation
\begin{eqnarray}
\theta^{\star}_{23} = \frac{\theta^{\star}_{12}
+ \theta^{\star}_{13}}{2} + \frac{\pi}{8} \; ,
\label{5}
\end{eqnarray}
just analogous to the conjecture made in Eq.~(\ref{2}). Here it is
worth mentioning that current experimental data on quark flavor
mixing and CP violation strongly indicates $\beta = \pi/8$ as a
good approximation for this smallest inner angle of the
$b$-quark-dominated unitarity triangle~\cite{Harrison:2025rkp,
Arkani-Hamed:2026uos}.

Taking $\theta^\star_{12}$, $\theta^\star_{13}$ and
$\theta^\star_{23}$ as the benchmark values of $\theta^{}_{12}$,
$\theta^{}_{13}$ and $\theta^{}_{23}$, we find that Eq.~(\ref{2})
can be easily derived from Eq.~(\ref{5}) if deviations of the
three neutrino mixing angles from their respective benchmarks
trivially satisfy the condition
\begin{eqnarray}
\theta^{\star}_{23} - \theta^{}_{23} =
\frac{\theta^{\star}_{12} - \theta^{}_{12}}{2} +
\frac{\theta^{\star}_{13} - \theta^{}_{13}}{2} \; .
\label{6}
\end{eqnarray}
With the help of the experimental data given in Eq.~(\ref{1})
and below Eq.~(\ref{2}), we arrive at
$\theta^{\star}_{12} - \theta^{}_{12}
\in \left[1.43^\circ, 1.86^\circ\right]$,
$\theta^{\star}_{13} - \theta^{}_{13}
\in \left[1.14^\circ, 1.38^\circ\right]$ and
$\theta^{\star}_{23} - \theta^{}_{23}
\in \left[0.23^\circ, 2.29^\circ\right]$ ({\rm or}
$\in \left[0.74^\circ, 2.52^\circ\right]$) at the $1\sigma$
level, implying that the three angle differences are all of
${\cal O}(1^\circ)$ in magnitude. So Eq.~(\ref{6}) has a very
good opportunity to be satisfied without any fine-tuning. This
observation assures that we may derive the angle relation in
Eq.~(\ref{2}) based on the magic angle relationship in Eq.~(\ref{5})
and the data-allowed condition in Eq.~(\ref{6}). Further precision
measurements, especially in the DUNE~\cite{DUNE:2015lol} and
Hyper-Kamiokande~\cite{Hyper-KamiokandeProto-:2015xww} long-baseline
accelerator-based neutrino oscillation experiments,
will provide stringent tests of Eqs.~(\ref{2}) and (\ref{6}).

We admit that a dynamical realization of Eq.~(\ref{6})
is expected to be strongly model-dependent. A simple but feasible
pathway to do so is that the neutrino sector exactly satisfies the
magic angle relation in Eq.~(\ref{5}) while the charged-lepton sector
contributes small corrections to lepton flavor mixing in a proper way.
A specific example of this kind will be taken in the next section.

One should also keep in mind that either Eq.~(\ref{2}) or Eq.~(\ref{6})
has been assumed to hold at the electroweak scale. Quantum
corrections to the three neutrino mixing angles are certainly
possible to spoil such a relationship~\cite{Xing:2019vks}, but the
relevant effects are usually small in the framework of the standard
electroweak interactions and thus may easily be contaminated
or even submerged by their respective experimental uncertainties.

\section{Expansion of flavor mixing}

We proceed to consider a special but still reasonable case of the
condition in Eq.~(\ref{6}), namely
\begin{eqnarray}
\theta^{\star}_{12} - \theta^{}_{12} =
\theta^{\star}_{13} - \theta^{}_{13} =
\theta^{\star}_{23} - \theta^{}_{23} \equiv \xi \; ,
\label{7}
\end{eqnarray}
which certainly allows us to obtain Eq.~(\ref{2}). In this
situation the standard parametrization of the $3\times 3$ unitary
lepton flavor mixing matrix $U$ as advocated by the Particle Data
Group~\cite{ParticleDataGroup:2026} can be expanded
in power of the small parameter $\xi$ as follows
\footnote{For simplicity, we have omitted the Majorana phases
of $U$ which are completely insensitive to the lepton-flavor-violating
but lepton-number-conserving neutrino oscillations.}:
\begin{eqnarray}
U = U^{}_\star + U^{}_1 \xi + \cdots + U^{}_n \xi^n + \cdots \; ,
\label{8}
\end{eqnarray}
where the leading-order term $U^{}_\star$ is given by
{\small
\begin{eqnarray}
U^{}_\star \hspace{-0.3cm} & = & \hspace{-0.35cm} \left(\begin{matrix}
\frac{\sqrt{2} + 1}{3} & \frac{\sqrt{2} + 1}{3\sqrt{2}} &
\frac{\sqrt{2} - 1}{\sqrt 6} e^{-{\rm i} \delta} \cr\vspace{-0.3cm}\cr
- \frac{1}{\sqrt 6} - \frac{\sqrt{2} - 1}{3\sqrt{2}} e^{{\rm i} \delta}
& \frac{1}{\sqrt{3}} - \frac{\sqrt{2} - 1}{6} e^{{\rm i} \delta}
& \frac{\sqrt{2} + 1}{2\sqrt{3}} \cr\vspace{-0.3cm}\cr
\frac{1}{\sqrt 6} - \frac{\sqrt{2} - 1}{3\sqrt{2}} e^{{\rm i} \delta}
& -\frac{1}{\sqrt{3}} - \frac{\sqrt{2} - 1}{6} e^{{\rm i} \delta}
& \frac{\sqrt{2} + 1}{2\sqrt{3}} \cr
\end{matrix} \right) \hspace{0.45cm}
\label{9}
\end{eqnarray}}
\hspace{-0.165cm}
with $\delta$ being the Dirac phase responsible for CP violation
in neutrino oscillations, and the elements of $U^{}_1$ are
{\small
\begin{eqnarray}
&& \left[U^{}_1\right]^{}_{e1} = \frac{1}{\sqrt{2}} \; ,
\quad \left[U^{}_1\right]^{}_{e2} = - \frac{1}{\sqrt{2}} \; ,
\nonumber \\
&& \left[U^{}_1\right]^{}_{e3} = -\frac{\sqrt{2} + 1}{\sqrt 6}
e^{-{\rm i} \delta} \; ,
\nonumber \\
&& \left[U^{}_1\right]^{}_{\mu 1} =
\frac{\sqrt{2} - 1}{\sqrt 6} + \frac{5 - \sqrt{2}}{6} e^{{\rm i}
\delta} \; ,
\nonumber \\
&& \left[U^{}_1\right]^{}_{\mu 2} =
\frac{\sqrt{2} + 1}{\sqrt{6}} + \frac{\sqrt{2} + 1}{3\sqrt{2}}
e^{{\rm i} \delta} \; ,
\nonumber \\
&& \left[U^{}_1\right]^{}_{\tau 1} =
-\frac{\sqrt{2} + 1}{\sqrt 6} + \frac{\sqrt{2} + 1}{6} e^{{\rm i}
\delta} \; ,
\nonumber \\
&& \left[U^{}_1\right]^{}_{\tau 2} =
\frac{\sqrt{2} - 1}{\sqrt{6}} + \frac{2\sqrt{2} - 1}{3\sqrt{2}}
e^{{\rm i} \delta} \; , \hspace{1cm}
\nonumber \\
&& \left[U^{}_1\right]^{}_{\mu 3} = -\frac{1}{\sqrt{3}} \; , \quad
\left[U^{}_1\right]^{}_{\tau 3} = \frac{\sqrt 2}{\sqrt{3}} \; .
\label{10}
\end{eqnarray}}
\hspace{-0.2cm} It is clear that $U^{}_\star$
itself is unitary and purely originates from the three magic
constant flavor mixing angles $\theta^\star_{12}$, $\theta^\star_{13}$
and $\theta^\star_{23}$, and hence it offers a benchmark basis for
the realistic lepton flavor mixing matrix $U$. On the other hand,
$\xi \sim 0.02$ serves as an effective measure of small breaking
effects of the $\mu$-$\tau$ reflection symmetry~\cite{Xing:2015fdg}.

Note that the unitarity of $U^{}_\star$ allows us to
rewrite $U$ as $U = U^\dagger_l U^{}_\nu$ with $U^{}_\nu = U^{}_\star$
from the neutrino sector and $U^{}_l = I + U^{}_\star U^\dagger_1 \xi
+ \cdots$ from the charged-lepton sector. This expression implies that
$-\xi$ measures a universal deviation of $\theta^{}_{12}$,
$\theta^{}_{13}$ and $\theta^{}_{23}$ from their respective magic
values, so as to offer a simple dynamical origin of Eq.~(\ref{7}).

Note also that the three flavor mixing angles
extracted from the expansion of $U$ in Eq.~(\ref{8}) always satisfy
their interrelation that has been conjectured in Eq.~(\ref{2}), simply
because Eq.~(\ref{7}) as a prerequisite used for this expansion does
not spoil Eq.~(\ref{2}) no matter whether the higher-order terms of
$U$ are taken into account or not. But the explicit forms of
$U^{}_n \xi^n$ (for $n = 1, 2, \cdots$) in the expansion of $U$
may help a lot for potential model building.

We find that it is good enough to keep the next-to-leading-order terms
of $U$ for fitting current neutrino oscillation data. In particular,
the Jarlskog invariant of leptonic CP violation~\cite{Jarlskog:1985ht,
Wu:1985ea} is obtained as
\begin{eqnarray}
{\cal J}^{}_\nu = \frac{\sqrt{2} + 1}{36\sqrt{3}} \left(1 -
\frac{13\sqrt{2} - 6}{2} \xi\right) \sin\delta \; ,
\label{11}
\end{eqnarray}
whose leading term is consistent with the one obtained in
Refs.~\cite{Xing:2010pn,Huang:2024xiq}. The assumption made in
Eq.~(\ref{7}) and its interesting consequences achieved in
Eqs.~(\ref{8})---(\ref{11}) will soon be experimentally tested.

\section{Summary}

In view of current neutrino oscillation data, especially those
from the JUNO and Daya Bay precision measurements, we have
conjectured a simple interrelation of three neutrino mixing angles:
$\theta^{}_{23} = \left(\theta^{}_{12} + \theta^{}_{13}\right)/2
+ \pi/8$. Its prediction
$\theta^{}_{23} = 43.46^{+0.51^\circ}_{-0.52^\circ}$
is compatible with the global-fit result of $\theta^{}_{23}$ to
a remarkably good degree of accuracy in the normal neutrino mass
ordering case. We have proposed a simple ansatz to derive this
phenomenologically favored and experimentally testable angle
relation by starting from a magic constant angle basis which
is closely associated with the well-known democratic and tribimaximal
flavor mixing patterns. A new expansion of the lepton flavor mixing
matrix in powers of a single small parameter has also been presented
along this line of thought, as a by-product of the proposed ansatz.	

The next-generation accelerator-based long-baseline neutrino
oscillation experiments are expected to offer stringent tests of the
interrelation of three neutrino mixing angles discussed in this work
in the near future. Provided this novel angle relation turns out to
survive such experimental tests, further efforts should be
made towards a complete model building based on some proper flavor
symmetry groups.

\section*{Acknowledgments}

The author is indebted to Zhibing Li for his kind encouragement
and to Wei Su for his warm hospitality at the Shenzhen Campus of
Sun Yat-sen University, where this ``flight" paper was finalized.
He is also grateful to Ibragim Alikhanov, Zi-Qiang Chen and Di Zhang
for kindly pointing out a few typos in the original preprint.

\section*{Funding}

This work was supported in part by the National Natural Science Foundation of
of China under Grant No. 12535007, and by the Scientific and Technological
Innovation Program of the Institute of High Energy Physics under Grant
No. E55457U2.

\section*{Data Availability}

No data and code supporting were used for the findings of this study.

\section*{Conflict of Interest}

The author declares no competing interests.


\begin{thebibliography}{99}

\bibitem{JUNO:2025gmd}
A.~Abusleme \textit{et al.} [JUNO],
``Measurement of reactor neutrino oscillation with the first JUNO data,''
Nature \textbf{654} (2026) no.8118, 343-348
[arXiv:2511.14593 [hep-ex]].

\bibitem{JUNO:2026tap}
T.~Adam \textit{et al.} [JUNO],
``Measurement of Neutrino Oscillation Parameters at JUNO and Indication of
the Neutrino Mass Ordering,''
[arXiv:2609.37895 [hep-ex]].

\bibitem{Alikhanov:2026jmw}
I.~Alikhanov,
``Another relation among the neutrino mass-squared differences?,''
[arXiv:2601.18781 [hep-ph]].

\bibitem{Alikhanov:2026bhi}
I.~Alikhanov,
``Novel dependence between neutrino mass splittings strongly supported by
initial JUNO results,''
[arXiv:2610.06738 [hep-ph]].

\bibitem{DayaBay:2022orm}
F.~P.~An \textit{et al.} [Daya Bay],
``Precision Measurement of Reactor Antineutrino Oscillation at
Kilometer-Scale Baselines by Daya Bay,''
Phys. Rev. Lett. \textbf{130}, no.16, 161802 (2023)
[arXiv:2211.14988 [hep-ex]].

\bibitem{Capozzi:2025wyn}
F.~Capozzi, W.~Giar{\`e}, E.~Lisi, A.~Marrone, A.~Melchiorri and A.~Palazzo,
``Neutrino masses and mixing: Entering the era of subpercent precision,''
Phys. Rev. D \textbf{111} (2025) no.9, 093006
[arXiv:2503.07752 [hep-ph]].

\bibitem{Esteban:2024eli}
I.~Esteban, M.~C.~Gonzalez-Garcia, M.~Maltoni, I.~Martinez-Soler, J.~P.~Pinheiro
and T.~Schwetz,
``NuFit-6.1: updated global analysis of three-flavor neutrino oscillations,''
JHEP \textbf{12} (2024), 216
[arXiv:2410.05380 [hep-ph]].

\bibitem{Fritzsch:1995dj}
H.~Fritzsch and Z.~Z.~Xing,
``Lepton mass hierarchy and neutrino oscillations,''
Phys. Lett. B \textbf{372} (1996), 265-270
[arXiv:hep-ph/9509389 [hep-ph]].

\bibitem{Fritzsch:1998xs}
H.~Fritzsch and Z.~z.~Xing,
``Large leptonic flavor mixing and the mass spectrum of leptons,''
Phys. Lett. B \textbf{440} (1998), 313-318
[arXiv:hep-ph/9808272 [hep-ph]].

\bibitem{Fritzsch:1999ee}
H.~Fritzsch and Z.~z.~Xing,
``Mass and flavor mixing schemes of quarks and leptons,''
Prog. Part. Nucl. Phys. \textbf{45} (2000), 1-81
[arXiv:hep-ph/9912358 [hep-ph]].

\bibitem{Harrison:2002er}
P.~F.~Harrison, D.~H.~Perkins and W.~G.~Scott,
``Tri-bimaximal mixing and the neutrino oscillation data,''
Phys. Lett. B \textbf{530} (2002), 167
[arXiv:hep-ph/0202074 [hep-ph]].

\bibitem{Xing:2002sw}
Z.~z.~Xing,
``Nearly tribimaximal neutrino mixing and CP violation,''
Phys. Lett. B \textbf{533} (2002), 85-93
[arXiv:hep-ph/0204049 [hep-ph]].

\bibitem{He:2003rm}
X.~G.~He and A.~Zee,
``Some simple mixing and mass matrices for neutrinos,''
Phys. Lett. B \textbf{560} (2003), 87-90
[arXiv:hep-ph/0301092 [hep-ph]].

\bibitem{Xing:2019vks}
Z.~z.~Xing,
``Flavor structures of charged fermions and massive neutrinos,''
Phys. Rept. \textbf{854} (2020), 1-147
[arXiv:1909.09610 [hep-ph]].

\bibitem{Feruglio:2019ybq}
F.~Feruglio and A.~Romanino,
``Lepton flavor symmetries,''
Rev. Mod. Phys. \textbf{93} (2021) no.1, 015007
[arXiv:1912.06028 [hep-ph]].

\bibitem{Ding:2024ozt}
G.~J.~Ding and J.~W.~F.~Valle,
``The symmetry approach to quark and lepton masses and mixing,''
Phys. Rept. \textbf{1109} (2025), 1-105
[arXiv:2402.16963 [hep-ph]].

\bibitem{Xing:2010pn}
Z.~z.~Xing,
``A Shift from Democratic to Tri-bimaximal Neutrino Mixing with
Relatively Large $\theta^{}_{13}$,''
Phys. Lett. B \textbf{696} (2011), 232-236
[arXiv:1011.2954 [hep-ph]].

\bibitem{Harrison:2025rkp}
P.~F.~Harrison and W.~G.~Scott,
``Unitarity triangle angles explained: a predictive new quark mass
matrix texture,''
JHEP \textbf{07} (2025), 155
[arXiv:2501.18508 [hep-ph]].

\bibitem{Arkani-Hamed:2026uos}
N.~Arkani-Hamed, C.~Figueiredo, L.~J.~Hall and C.~A.~Manzari,
``The Very Nearly Right Theory of Flavor,''
[arXiv:2607.27315 [hep-ph]].

\bibitem{DUNE:2015lol}
R.~Acciarri \textit{et al.} [DUNE],
``Long-Baseline Neutrino Facility (LBNF) and Deep Underground Neutrino
Experiment (DUNE):
Conceptual Design Report, Volume 2: The Physics Program for DUNE at LBNF,''
[arXiv:1512.06148 [physics.ins-det]].

\bibitem{Hyper-KamiokandeProto-:2015xww}
K.~Abe \textit{et al.} [Hyper-Kamiokande Proto-],
``Physics potential of a long-baseline neutrino oscillation experiment using
a J-PARC neutrino beam and Hyper-Kamiokande,''
PTEP \textbf{2015} (2015), 053C02
[arXiv:1502.05199 [hep-ex]].

\bibitem{ParticleDataGroup:2026}
F.~Takahashi \textit{et al.} [Particle Data Group],
``Review of particle physics,''
Int. J. Mod. Phys. A \textbf{41} (2026) 2630011

\bibitem{Xing:2015fdg}
Z.~z.~Xing and Z.~h.~Zhao,
``A review of $\mu$-$\tau$ flavor symmetry in neutrino physics,''
Rept. Prog. Phys. \textbf{79} (2016) no.7, 076201
[arXiv:1512.04207 [hep-ph]].

\bibitem{Jarlskog:1985ht}
C.~Jarlskog,
``Commutator of the Quark Mass Matrices in the Standard Electroweak Model
and a Measure of Maximal CP Nonconservation,''
Phys. Rev. Lett. \textbf{55} (1985), 1039

\bibitem{Wu:1985ea}
D.~d.~Wu,
``The Rephasing Invariants and CP,''
Phys. Rev. D \textbf{33} (1986), 860

\bibitem{Huang:2024xiq}
J.~Huang,
``A nearest-neighbor expansion of lepton flavor mixing in powers of the
$\mu$-$\tau$ permutation symmetry breaking effect,''
Phys. Lett. B \textbf{856} (2024), 138898
[arXiv:2405.20871 [hep-ph]].

\end{thebibliography}
\end{document}